\documentclass{article}
\usepackage[utf8]{inputenc}
\usepackage{graphicx}
\usepackage{amsmath}
\usepackage{amsfonts}
\usepackage{amssymb}
\usepackage{caption}
\usepackage{subcaption}
\usepackage{cite}
\usepackage{geometry}
\usepackage{lipsum}
\usepackage[english]{babel}  
\usepackage{authblk}
\usepackage{chemformula}
\usepackage{booktabs}
\usepackage{float} 

\title{New opportunities for high pressure hydrogen achieved by fullerane vibrating modes: an ab initio study}
\author[1]{Leonard Constantin Gebac\thanks{Corresponding author: \texttt{leonard.gebac@unibuc.ro}}}
\author[1]{Vasile Bercu}

\affil[1]{Faculty of Physics, University of Bucharest, 405 Atomistilor Street, Magurele, Romania}

\date{\today}

\begin{document}

\maketitle

\begin{abstract}
    The encapsulation of hydrogen within fullerene/fullerane cages offers a promising avenue for studying high pressure hydrogen dynamics. Through ab initio molecular dynamics simulations, we investigate the behavior of a system consisting of hydrogen atoms enclosed in a \ch{C20H20} dodecahedrane. Our findings reveal significant structural and dynamical changes as the cage undergoes compression, corresponding to radial symmetric vibration. We analyze geometric, energetic, and thermodynamic parameters, highlighting correlations and observing behavior analogous to high pressure phases of hydrogen. Notably, our study bridges the gap between theory and experiment by proposing a novel approach to achieving high pressures and temperatures experimentally. These results not only contribute to the understanding of hydrogen behavior under extreme conditions but also hold implications for the quest to attain metallic hydrogen—a milestone in materials science with potential applications in various fields.
\end{abstract}
\noindent\textbf{Keywords:} High pressure hydrogen dynamics, Hydrogen encapsulation, Radial symmetric vibration, Fullerane, Ab initio molecular dynamics

\section{Introduction}
\label{introduction}
Hydrogen, the simplest and most abundant element in the universe, exhibits a rich array of physical behaviors under extreme conditions, particularly at high pressures. At standard ambient conditions, hydrogen predominantly exists in molecular form (\ch{H2}), with each molecule comprising two hydrogen atoms bonded by a stable covalent bond. However, as pressure increases, hydrogen undergoes dramatic transformations, transitioning through distinct phases with unique properties. Notably, the prospect of metallic hydrogen has captivated scientists for decades, holding implications for various fields ranging from energy production to planetary science \cite{guillot2005interiors}. Academician Vitaly Ginzburg, in his seminal list of significant challenges in physics for the 21st century, ranked metallic hydrogen as the third most pressing challenge \cite{ginzburg2001problems}.

Experimental observations have revealed a series of distinct phases that hydrogen undergoes under varying pressure and temperature conditions. Under large values of pressures, these include Solid Phase I, characterized by a hexagonal close-packed structure \cite{phase1, review_metallicH_1}, Phase II \cite{liu2017high}, known as the "broken-symmetry" phase \cite{silvera1981new}, Phase III \cite{hemley1988phase, howie2012proton, howie2012mixed, eremets2019semimetallic, ji2019ultrahigh, akahama2010evidence} with its hcp structure and intense infrared activity \cite{hanfland1993novel}, Phase IV marked by alternating layers of six-atom rings and free molecules \cite{howie2012mixed, pickard2007structure, ji2019ultrahigh}, and Phase V, a precursor to full atomic and metallic states \cite{dalladay2016evidence}. Recent experimental evidence has also suggested the existence of Phase VI \cite{loubeyre2020synchrotron, monacelli2023quantum} at extremely high pressures. At even greater pressures, a study claimed to have synthesized metallic hydrogen based on reflectivity measurements at 495 GPa \cite{dias2017observation}, in a good agreement with theoretical predictions\cite{borinaga2018strong}, but their findings were questioned among the high pressure hydrogen scientific community \cite{goncharov2017comment, silvera2017response, debate_hydrogen}. Additionally, hydrogen manifests in two distinct liquid phases: molecular liquid and metallic liquid. The molecular liquid phase occurs at moderate pressures and temperatures and is characterized by the persistence of hydrogen molecules in a fluid state, retaining their molecular identity \cite{liquidH}. Conversely, at higher pressures and temperatures, hydrogen undergoes a transition to a metallic liquid phase, where the hydrogen atoms lose their molecular identity \cite{silvera2017metallic, morales2010pnas}.

Nevertheless, uncertainties persist, particularly regarding pressure measurements at such extreme conditions. The need for pressure rescaling has been emphasized in recent studies, acknowledging potential discrepancies of up to -20\% at 500 GPa \cite{eremets2023universal}.

Considering these experimental results accompanied by uncertainties and even debates, numerical calculations and simulations become essential tools in predicting the structures of the observed phases of solid hydrogen. Previous studies \cite{nagara1992stable, pickard2007structure, pickard2012density, monserrat2016hexagonal, monserrat2018structure} have utilized density functional theory (DFT) and structure search algorithms but have faced difficulties in distinguishing the most stable structures due to small differences in enthalpy values and the choice of exchange-correlation functionals \cite{azadi2017role, azadi2019unconventional, azadi2013fate, yang2020first}.

Structure search methods show promise in identifying phases at low temperatures with harmonic phonons, but hydrogen's behavior, characterized by pronounced quantum anharmonic effects, even at room temperature, poses challenges for these algorithms \cite{singh2014anharmonicity, liu2013proton, min1986pressure}. From this perspective, ab initio molecular dynamics (AIMD) simulations emerge as a suitable approach for exploring this region, contributing to the understanding of the symmetry and the structure at both low and high-temperatures \cite{liu2012roomtempstruct}. In general, ab initio studies are extremely valuable in examining material properties under high pressure conditions, as demonstrated in several recent studies \cite{BSi_abinitio, Sr2GeN2_abinitio, ethanol_exp_abinitio, Vn_abinitio_exp}. More sophisticated methods \cite{monacelli2023quantum, drummond2015quantum, azadi2013quantum, azadi2014dissociation, geng2017prediction, helled2020understanding, chen2014roomtempPD}, such as quantum Monte Carlo (QMC) and path integral molecular dynamics (PIMD), are used to accurately map the phase diagram of solid hydrogen, taking into account quantum anharmonic effects. However, discrepancies still exist between different calculation methods \cite{li2024high}.

All the computational methods outlined above, whether based on structure search methods, which are static, following the isothermal-isobaric (NPT) ensemble where pressure is computationally controlled, or molecular dynamics simulations, which are non-static, either ab initio, quantum Monte Carlo, or path integral based, have one thing in common: they seek a structure following computationally imposed pressure or temperature.

In this context, our study focuses on AIMD simulations in which the pressures and temperatures are dynamically induced by a \ch{C20H20} molecular cage which encapsulates hydrogen atoms. The dodecahedrane was firstly synthesized in 1983 \cite{c20h20_sinteza} and is known for it's stability, even when incorporating noble gas atoms \cite{he_inside_c20h20, he_C20H20_eu, ne_C20H20_eu}. We investigate the influence of the radial symmetric vibration mode of the \ch{C20H20} molecule on the encapsulated hydrogen atoms. Under the immense pressure exerted by the cage, reaching hundreds of gigapascals, we observe a transformation in the behavior of the hydrogen atoms initially grouped as three hydrogen molecules. By utilizing the radial breathing mode of dodecahedrane, we aim to bridge the gap between theory and experiment. Our approach offers insights into the dynamic behavior of hydrogen under extreme pressures. Moreover, we propose a novel method to achieve these extreme pressures and high temperatures experimentally. This method could lead to the formation of metallic hydrogen, whether in its liquid or solid phase. Our study is meant to prove that this goal can be achieved by inducing a symmetric normal mode of vibration in a molecular cage which sequestrates hydrogen atoms. While our investigation acknowledges the potential for further extensions to larger cages (such as \ch{C60} or carbon nanotubes), the primary focus of this study is to establish the feasibility of our proposed method. We have chosen to use only six hydrogen atoms enclosed within a \ch{C20H20} molecular cage as a demonstrative model. This choice emphasizes the method rather than the number of atoms. Additionally, it reflects findings from other studies, which indicate that hydrogen atoms can adopt a hexagonal close-packed symmetry. This configuration effectively replicates that symmetry. Our simulations reveal intricate correlations between the cage radius, the difference between lowest unoccupied molecular orbital (LUMO) and highest occupied molecular orbital (HOMO) energies, internal \ch{H-H} distances, and the pressure experienced by the \ch{H6} atomic system. These findings not only enhance our understanding of high pressure hydrogen dynamics but also offer valuable insights into the quest for metallic hydrogen and its implications across multiple scientific disciplines, in particularly in the hydrogen energy field and sustainable energy \cite{H_storage_review_1, H_storage_cage_water_enf, H_storage_MD_1, H_diff_storg_enf_1, H_diff_storg_enf_2, H_diff_storg_enf_3,H_diff_storg_enf_4, H_storage_G_CNT_MD_1, SHAKIL2024740, KHALILOV2024604, OLIVER2024540, ZHANG202323909, WU201214211, HARRISON20172223, FLAMINA2024119657, TIAN2021968}.
While this work focuses on short-timescale pressure dynamics, future studies will investigate extended simulations to assess stability under sustained conditions.

\section{Theory and Methods}
\label{methods}
\textit{Ab initio} molecular dynamics simulations were performed using the TeraChem code \cite{ufimtsev2009quantum}. The basis set employed is aug-pcseg-1, and the chosen functional is rCAM-B3LYP \cite{cohen2007development}, a functional corrected for long-range interactions with an improved description of systems with fractional number of electrons. This functional has shown very good results compared to other functionals used in calculations where electronic transfer is important \cite{malek2014testing}. Here, it was used in the spin-unrestricted variant. During the simulations, charge transfer between the internal hydrogen atoms and the atoms forming the cage may occur; hence, this functional was chosen for its demonstrated capability in describing such systems accurately \cite{malek2014testing}. The spin-unrestricted variant was selected because, in molecular dynamics, given the initial conditions, the system will evolve over time with considerable variations in interatomic distances, and unrestricted methods provide a more accurate description of the physical behaviour of the system \cite{uhf_motivation}. The aug-pcseg-1 basis set was chosen as it is specifically designed to be used with density functional theory methods \cite{aug-pcseg-1_jensen}.

The molecular dynamics simulations were conducted under the microcanonical (NVE) ensemble with a time step of 0.5 fs for a total duration of 500 fs (1000 steps). The extended Lagrangian method \cite{reversible_d_integrator} was employed for motion equation integrator, with initial velocities set to zero and the \ch{C20H20} cage initialized in an expanded configuration (1.2× equilibrium dimensions). A grid of 2683 calculation points per atom was used, and geometric, energetic, and thermodynamic quantities were monitored throughout. The short simulation window (0.5 ps) was chosen to capture the critical pressure dynamics during the cage’s symmetric vibrational cycles, which induce transient metallic hydrogen signatures.

The input files for the MD simulation, the single-point energy calculation and the results obtained at each time-step are available online \cite{gebac_2025_14843440}.

\subsection{Geometric Parameters}

One of the geometric parameters studied was the cage radius. This parameter was chosen so that it would show the normal mode of vibration. For this, we used two geometric quantities. The first one is denoted as $R_{C_{20}}$ and is defined as the average distance between the system's center of mass and the carbon atoms comprising the dodecahedrane. The second one is defined in a similar way by using the average distance to the hydrogen atoms ($R_{H_{20}}$) of the dodecahedrane.

Another geometric parameter, $R_{H_{6}}$, represents the average distance between the system's center of mass and the encapsulated hydrogen atoms. Concurrently, we calculated the volume enclosed by the hydrogen atoms, denoted as $V_{H_{6}}$, and the equivalent volume of a sphere with a radius equal to $R_{H_{6}}$, denoted as $V_{R_{H_{6}}}$. By comparing these volumes, we can calculate their ratio, which provides valuable insights into the planarity of the instantaneous configuration of the six hydrogen atoms.

The volume $V_{H_{6}}$ was determined using a methodology akin to that described in reference \cite{zakirova2020}. Initially, the position of the mass center of the \ch{H6} system was computed. Subsequently, the total volume $V_{H_{6}}$ was calculated by summing the volumes of tetrahedra formed by the coordinates of every combination of three points, representing the positions of three hydrogen atoms, with each tetrahedron's apex positioned at the mass center. The volume of a tetrahedron is given by:

\begin{eqnarray}\label{vol_tetraedru}
	\begin{aligned}
		V_{123} = \frac{1}{6}|\vec{v_1}\cdot(\vec{v_2}\times\vec{v_3})|
	\end{aligned}
\end{eqnarray}
Where the vectors $\vec{v_i}$, are defined by the position differences:

\begin{eqnarray}\label{vec_123}
	\begin{aligned}
		\vec{v_i} = \vec{r_i} - \vec{r_{0}},\ i = \overline{1,3}
	\end{aligned}
\end{eqnarray}
Utilizing $\vec{r_{i}}$ to represent the positions of three points labeled generically as 1, 2, and 3, and $\vec{r_{0}}$ to denote the position of the center of mass of \ch{H6}, the volume $V_{H_{6}}$ is computed as:

\begin{eqnarray}\label{vol_H6}
	\begin{aligned}
		V_{H_{6}} = \sum_{i=1}^{4}\sum_{j=i+1}^{5}\sum_{k=j+1}^{6}V_{ijk}
	\end{aligned}
\end{eqnarray}

\subsection{Energetic Parameters}

Energetic parameters were defined based on the conservation equation of the total energy:

\begin{eqnarray}\label{cons_en_total_H6}
	\begin{aligned}
		K_{C_{20}H_{20}}(t) + K_{H_{6}}(t) + Q_{H_{6}@C_{20}H_{20}}(t) = const.
	\end{aligned}
\end{eqnarray}

Where $K_{C_{20}H_{20}}(t)$ represents the total kinetic energy of the fullerane, $K_{H_{6}}(t)$ represents the total kinetic energy of the system formed by the six encapsulated hydrogen atoms, and $Q_{H_{6}@C_{20}H_{20}}(t)$ represents the total potential energy of the \ch{H6@C20H20} system. $Q_{H_{6}@C_{20}H_{20}}(t)$ can be expressed as:

\begin{eqnarray}\label{en_pot_total_H6}
	\begin{aligned}
		Q_{H_{6}@C_{20}H_{20}}(t) = Q_{H_{6}}(t) + Q_{C_{20}H_{20}}(t) + U_{H_{6}-C_{20}H_{20}}(t)
	\end{aligned}
\end{eqnarray}

Where $Q_{H_{6}}(t)$ represents the total potential energy of the system formed by the six hydrogen atoms, $Q_{C_{20}H_{20}}(t)$ denotes the total potential energy of the dodecahedrane, and $U_{H_{6}-C_{20}H_{20}}(t)$ represents the interaction potential energy between \ch{H6} and \ch{C20H20}. $Q_{H_{6}}$ and $Q_{C_{20}H_{20}}$ are calculated separately at each time step of the simulation. Thus, $Q_{H_{6}}(t)$ takes into account, at each time step, the deformation of the hydrogen system, which is induced by the supramolecular cage.

Additionally, the energies of the HOMO (highest occupied molecular orbital) and LUMO (lowest unoccupied molecular orbital), as well as the bandgap were monitored ($\Delta_{LUMO,HOMO} = E_{LUMO} - E_{HOMO}$). This energy levels correspond to the entire system \ch{H6@C20H20}.

\subsection{Thermodynamic Parameters}

In addition to the geometric and energetic parameters, two thermodynamic physical quantities were also observed and analyzed: pressure and temperature. The pressure is associated with the encapsulated atomic system evolving inside the fullerane, referring to the pressure experienced by the internal hydrogen atoms induced by the \ch{C20H20} molecule. It is calculated using the following equation \cite{bader_pressure, liao2019comparative}:

\begin{eqnarray}\label{def_pres}
	\begin{aligned}
		P = -\frac{\partial U}{\partial V}
	\end{aligned}
\end{eqnarray}
Numerically, the pressure is computed as the negative ratio of the variation of the internal energy of the system of interest and the variation of its volume. In our case, the internal energy is identical to the sum between the total potential energy of the system, $Q_{H_{6}}[i]$, and it's kinetic energy, $K_{H_{6}}[i]$, calculated at each time step $i$. Thus, at each time step:

\begin{eqnarray}\label{def_num_pres}
	\begin{aligned}
		P[i] = -\frac{Q_{H_{6}}[i] + K_{H_{6}}[i] - Q_{H_{6}}^0}{V_{H_{6}}[i] - V_{H_{6}}^0}
	\end{aligned}
\end{eqnarray}
Here, $Q_{H_{6}}^0$ and $V_{H_{6}}^0$ denote the total potential energy and volume, respectively, of the \ch{H6} structure following the geometry optimization procedure \cite{terachem_GO_new} applied to this system. The positions of the hydrogen atoms were geometry optimized, starting from the initial configuration at time step 0. The final optimized configuration was planar, with the hydrogen atoms positioned at considerable distances from each other. In order to be able to apply Equation \ref{def_num_pres}, an intermediate configuration was selected from which variations in the total system energy were negligible (approximately 0.01 \%). At this point, the values $Q_{H_{6}}^0$ and $V_{H_{6}}^0$ were computed, enabling the calculation of pressure at each time step. Moreover, $K_{H_{6}}^0$ is considered equal to zero and is omitted from the equation \ref{def_num_pres}.

The temperature associated to the \ch{H6} system is defined in a classical fashion by the average kinetic energy of the six encapsulated hydrogen atoms \cite{temp_def}:

\begin{eqnarray}\label{temperature}
	\begin{aligned}
		T[i] = \frac{2}{3}\frac{<K_{H_{j}}[i]>}{k_B},\ j=\overline{1,6}
	\end{aligned}
\end{eqnarray}

\section{Results and Discussion}
\label{sec:results}
For enhanced clarity in illustrating the correlations between the aforementioned physical parameters, they have been graphically integrated. While the total simulation duration spans 0.5 ps (1000 time steps), only half of this time range is represented to facilitate a more detailed observation of the dependencies and correlations among each quantity.

\begin{figure}[h!]
	\centering
	\includegraphics[width=0.8\textwidth]{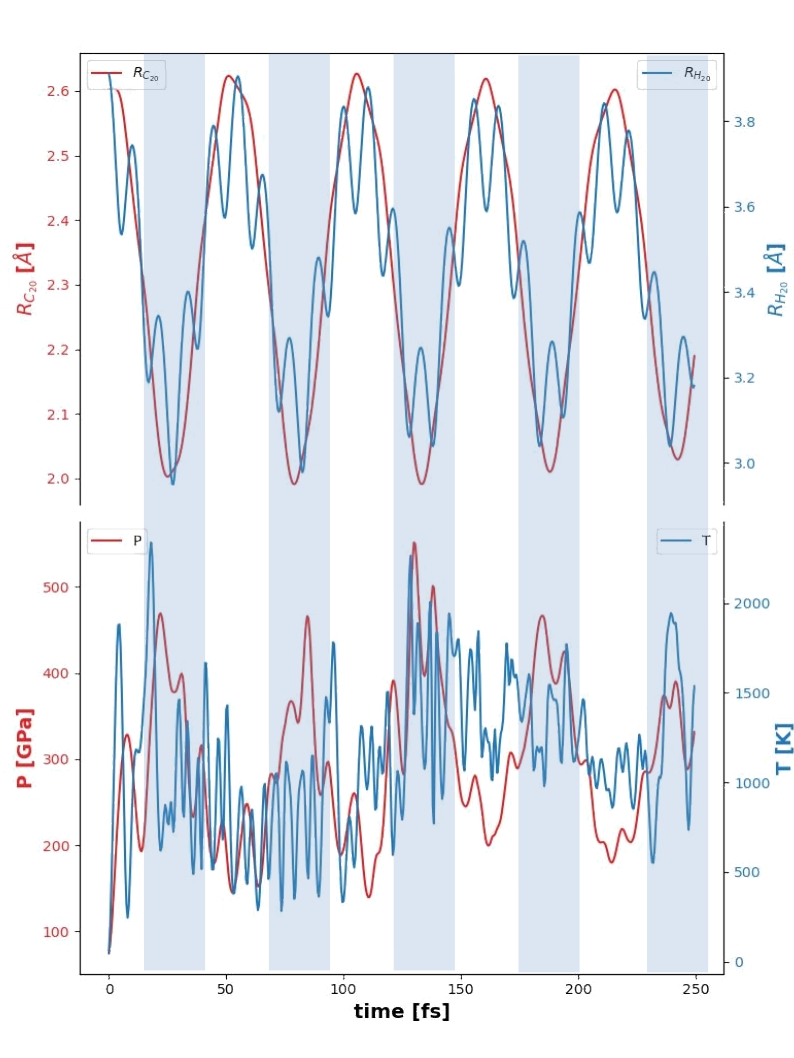}
	\caption{\textit{\small Top: Evolution of the cage radius, $R_{C_{20}}$ (red) and $R_{H_{20}}$ (blue). Bottom: Evolution of the pressure experienced by the \ch{H6} system (red) and the temperature (blue). Certain areas are colored to highlight correlations between the physical quantities.}}
	\label{R_C20_R_H20_P_delU_m1}
\end{figure}

In Fig. \ref{R_C20_R_H20_P_delU_m1}, the top section illustrates the radii of the dodecahedrane, with $R_{C_{20}}$ in red and $R_{H_{20}}$ in blue. It can be observed that the cage begins to compress from the initial time step, as its starting size exceeds its equilibrium size by a factor of $k=1.2$. As the system naturally evolves towards a more energetically stable configuration, the cage compresses, resulting in a decrease in $R_{C_{20}}$. Correspondingly, the radius $R_{H_{20}}$ also decreases. Notably, within the time interval where $R_{C_{20}}$ completes one oscillation, $R_{H_{20}}$ undergoes approximately five variations. This is understandable given that the variation in $R_{H_{20}}$ is influenced by the characteristic vibration frequency of the \ch{C-H} bond, which is higher than that of the \ch{C-C} bond, dictating the variation frequency of $R_{C_{20}}$. A clear periodicity of this oscillation type is observed over time.

In the bottom section of Fig. \ref{R_C20_R_H20_P_delU_m1}, the evolution of both pressure and temperature of the encapsulated \ch{H6} system over time is shown. The colored portions of the graph highlight the correlations between the quantities represented in the top and bottom sections. It is notable that pressure peaks align with the minimum values of $R_{C_{20}}$. During these instances of cage compression, the highest pressures within the \ch{H6} system are induced, reaching between 400 and 500 GPa. Additionally, except for the initial four time steps, the pressure does not fall below 100 GPa. As depicted in Fig. \ref{R_C20_R_H20_P_delU_m1}, the temperature ranges between 200 and 2500 K.

\begin{figure}[h!]
	\centering
	\includegraphics[width=0.8\textwidth]{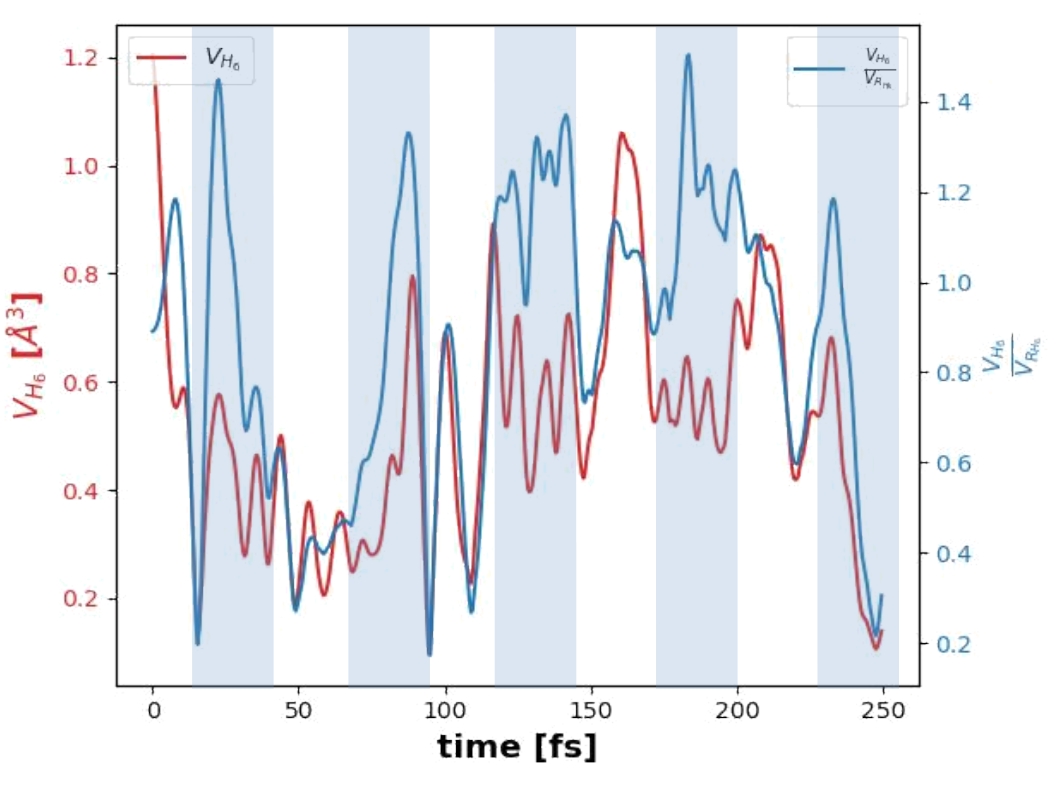}
	\caption{\textit{\small Evolution of the volume determined by the 6 hydrogen atoms (red) and the ratio between this volume and the equivalent volume of a sphere with a radius equal to the average distance between the center of mass and the 6 hydrogen atoms (blue). Certain areas are colored to highlight correlations between the analyzed and represented physical quantities.}}
	\label{raport_V_H6_m3}
\end{figure}

In Fig. \ref{raport_V_H6_m3}, the time evolution of the volume determined by the six encapsulated hydrogen atoms, $V_{H_{6}}$ (in red), and the ratio between this volume and the equivalent volume of a sphere with a radius equal to the average distance between the center of mass and the six \ch{H} atoms, $V_{R_{H_{6}}}$, is depicted. This representation is instrumental in qualitatively assessing the geometric configuration of the six atoms at various moments in time. Essentially, the system transitions through three well-defined configurations with transitional geometric distributions in between.

The first configuration is three-dimensional, characterized by large volume ratios (the peaks of the blue curve in Fig. \ref{raport_V_H6_m3}) and small volumes $V_{H_{6}}$. Another encountered configuration is the quasi-molecular one, which features both large volume ratios and large $V_{H_{6}}$ volumes. The third configuration is the planar-hexagonal one, characterized by very small $V_{H_{6}}$ volumes and correspondingly small volume ratios. In Fig. \ref{raport_V_H6_m3}, these configurations are identifiable where both $V_{H_{6}}$ and $V_{R_{H_{6}}}$ values are small and overlap.

\begin{figure}[h!]
	\centering
	\includegraphics[width=0.8\textwidth]{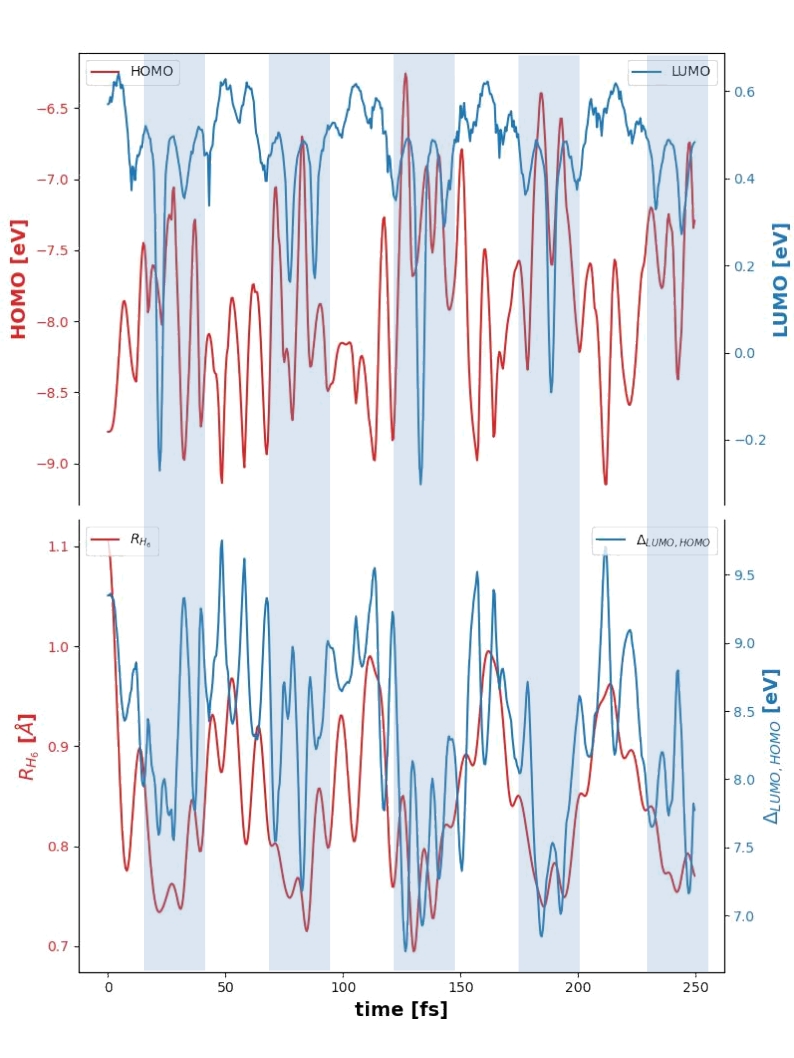}
	\caption{\textit{\small Top: Evolution of HOMO (red) and LUMO (blue) energy levels. Bottom: Evolution of the average distance between the center of mass and the 6 hydrogen atoms, $R_{H_6}$ (red), and the difference between the two energy levels LUMO - HOMO (blue). Certain areas are colored to highlight correlations between the analyzed and represented physical quantities.}}
	\label{H_L_R_H6_delHL_m1}
\end{figure}

In Fig. \ref{H_L_R_H6_delHL_m1}, the upper section depicts the variations in energy levels, with the HOMO represented in red and the LUMO in blue. These variations correspond to the entire \ch{H6@C20H20} system. The shaded vertical bands, as indicated in Fig. \ref{R_C20_R_H20_P_delU_m1}, highlight the periods of compression induced by the reduction in the radius $R_{C_{20}}$. During these compression moments, the HOMO level tends to increase compared to the extension periods, while the LUMO level decreases, reaching its minima. This behavior is further confirmed in the lower section of the Fig., which illustrates the band gap in blue. The band gap narrows to approximately 7 eV during compression, compared to about 9.5 eV during extension, a behavior which was anticipated based on previous studies, for example, in reference \cite{bandgap_pressure}, is shown that for dense hydrogen, the band gap decrease linearly with increasing pressure, from 10.9 to 6.57 eV. 

Additionally, $R_{H_6}$ is correlated with the pressure increase in the \ch{H6} system due to the dodecahedrane's compression (see bottom of Fig. \ref{R_C20_R_H20_P_delU_m1}). This distance follows the "breathing" mode of the dodecahedrane, showing a decrease in $R_{H_6}$ as the fullerane compresses, thereby compressing the six-atom system.

Figs. \ref{R_C20_R_H20_P_delU_m1}, \ref{raport_V_H6_m3}, and \ref{H_L_R_H6_delHL_m1} collectively elucidate the evolution of the \ch{H6@C20H20} system and the impact of a radially symmetric vibrational mode on the \ch{H6} assembly. The compression of the molecular cage induces significant compression within the \ch{H6} system, with pressures surpassing 300 GPa at peak compression points (see bottom of Fig. \ref{R_C20_R_H20_P_delU_m1}). The geometric configurations associated with these extreme pressures are distinctly three-dimensional, as indicated by the peaks in the volume ratio (see the blue curve of Fig. \ref{raport_V_H6_m3}). These high pressure configurations lead to notable changes in the system's energy levels. Additionally, as the cage compresses the \ch{H6} system, the band gap of the entire \ch{H6@C20H20} structure narrows, highlighting the dynamic interplay between structural and electronic properties under extreme conditions.

\begin{table}[]
	\centering
	\caption{Pressure and temperature characteristics taken from literature used to identify the analogues phases explored by the \ch{H6} system}
	\label{tab_phases}
	\begin{tabular}{@{}lll@{}}
		\toprule
		\textbf{Phase} & \textbf{Pressure (GPa)} & \textbf{Temperature (K)} \\ \midrule
		Molecular Liquid\cite{review_metallicH_1, liquidH} & 70-120 & 850-2000 \\
		& 120-140 & 800-1500 \\
		& 140-160 & 750-1300 \\
		& 160-180 & 650-1000 \\
		& 180-200 & 600-750 \\
		& 200-225 & 550-700 \\
		& 225-250 & 500-650 \\
		& 250-325 & 520-600 \\ \midrule
		Metallic Liquid\cite{morales2010pnas, silvera2017metallic} & 100-120 & $> 2000$ \\
		& 120-140 & $> 1500$ \\
		& 140-160 & $> 1300$ \\
		& 160-180 & $> 1000$ \\
		& 180-200 & $> 750$ \\
		& 200-225 & $> 700$ \\
		& 225-250 & $> 650$ \\
		& 250-325 & $> 600$ \\
		& 325-425 & $> 520$ \\
		& 425-500 & $> 500$ \\ 
		& $> 500$ & $> 500$ \\ \midrule
		Phase 1\cite{review_metallicH_1, phase1} & 70-120 & 100-850 \\
		& 120-140 & 100-800 \\
		& 140-160 & 100-750 \\
		& 160-180 & 200-650 \\
		& 180-200 & 280-600 \\
		& 200-225 & 350-550 \\
		& 225-250 & 450-500 \\ \midrule
		Phase II\cite{liu2017high} & 70-160 & 0-100 \\ \midrule
		Phase III\cite{hemley1988phase, howie2012proton, howie2012mixed, eremets2019semimetallic, ji2019ultrahigh, akahama2010evidence} & 160-180 & 0-200 \\
		& 180-200 & 0-280 \\
		& 200-225 & 0-350 \\
		& 225-250 & 0-300 \\
		& 250-325 & 0-250 \\
		& 325-425 & 0-200 \\ \midrule
		Phase IV\cite{howie2012mixed, pickard2007structure, ji2019ultrahigh} & 225-250 & 300-450 \\
		& 250-325 & 250-520 \\ \midrule
		Phase V\cite{dalladay2016evidence} & 325-425 & 200-520 \\ \midrule
		Phase VI\cite{loubeyre2020synchrotron, monacelli2023quantum} & 425-500 & 0-500 \\ \midrule
		Metallic Solid & $> 500$ & 0-500 \\ \bottomrule
	\end{tabular}
\end{table}

\begin{figure}[h!]
	\centering
	\includegraphics[width=0.8\textwidth]{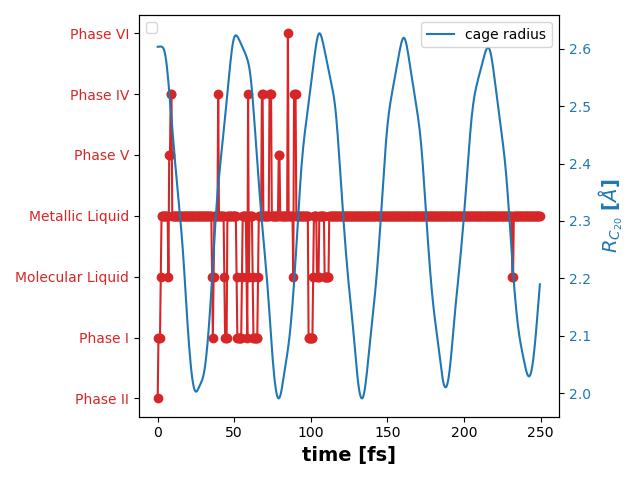}
	\caption{\textit{\small \ch{H6} system evolution across different analogues phases (red) and the dodecahedrane radius (blue) as a function of time.}}
	\label{R_C20_faze}
\end{figure}

At each time-step, the system's classical analogue to a thermodynamic phase, as described in Table \ref{tab_phases}, can be identified. For the sake of simplicity and ease of reference throughout the text, we will refer to this classical analogue to a thermodynamic phase simply as "phase".
Fig. \ref{R_C20_faze} illustrates the evolution of the six-hydrogen atom system, presenting the phases it traverses alongside the cage radius over time. Notably, during peak compression of the cage, the \ch{H6} system transitions into the metallic liquid analogue state, whereas during cage extension, it reverts to the molecular liquid analogue state. Between these extremes, the system exhibits transitions between these two analogues states.

\begin{table}[]
	\centering
	\caption{Mean values and standard deviations for each phase}
	\label{results}
	\small 
	\setlength{\tabcolsep}{1pt} 
	\begin{tabular}{|l|c|c|c|c|c|}
		\hline
		Phase & H-H Distance [\AA] & $V_{H_6}$ [\AA$^3$] & $\frac{V_{H_6}}{V_{R_{H_6}}}$ & P [GPa] & T [K] \\ \hline
		Molecular Liquid & 0.933 $\pm$ 0.075 & 0.496 $\pm$ 0.272 & 0.681 $\pm$ 0.336 & 210.066 $\pm$ 54.027 & 732.694 $\pm$ 212.700 \\ \hline
		Metallic Liquid & 0.904 $\pm$ 0.073 & 0.578 $\pm$ 0.231 & 0.927 $\pm$ 0.301 & 325.259 $\pm$ 88.800 & 1680.330 $\pm$ 693.512 \\ \hline
		Solid phase I & 0.966 $\pm$ 0.073 & 0.530 $\pm$ 0.257 & 0.613 $\pm$ 0.199 & 167.763 $\pm$ 35.556 & 454.792 $\pm$ 118.776 \\ \hline
		Solid phase II & 1.169 $\pm$ $-$ & 1.204 $\pm$ $-$ & 0.891 $\pm$ $-$ & 74.861 $\pm$ $-$ & 46.609 $\pm$ $-$ \\ \hline
		Solid phase IV* & 0.810 $\pm$ 0.007 & 0.259 $\pm$ 0.030 & 0.511 $\pm$ 0.076 & 286.927 $\pm$ 24.302 & 445.988 $\pm$ 78.162 \\ \hline
		Solid phase IV & 0.849 $\pm$ 0.052 & 0.441 $\pm$ 0.223 & 0.811 $\pm$ 0.345 & 290.483 $\pm$ 26.741 & 417.797 $\pm$ 74.687 \\ \hline
		Solid phase V & 0.833 $\pm$ 0.041 & 0.452 $\pm$ 0.102 & 0.992 $\pm$ 0.192 & 342.816 $\pm$ 14.690 & 327.101 $\pm$ 55.446 \\ \hline
		Solid phase VI & 0.775 $\pm$ $-$ & 0.445 $\pm$ $-$ & 1.201 $\pm$ $-$ & 464.146 $\pm$ $-$ & 463.448 $\pm$ $-$ \\ \hline
	\end{tabular}
\end{table}

Table \ref{results} outlines the primary outcomes of the simulation. Here, we introduce the 'H-H Distance' metric which denotes the average of the six shortest distances between hydrogen atoms. This calculation offers insights into the interatomic distances among neighboring hydrogens, particularly relevant when the \ch{H6} system adopts a hexagonal structure geometric configuration, allowing for comparisons with existing literature data.

Among the 1000 analyzed data points, only 12 align with phase IV as per the criteria delineated in Table \ref{tab_phases}. However, 7 of these points exhibit consistent numerical values for volumes and volume ratios. Consequently, these data points have been evaluated, denoted as IV*. Notably, during phase IV*, the average hydrogen-hydrogen distance measures 0.810 \AA\ $\pm$ 0.007, closely resembling the 0.82 \AA\ value suggested by reference \cite{howie2012mixed}. Additionally, minimal standard deviations associated with other parameters (volume and volume ratio) further validate the association of these points with phase IV.

For phases II and VI, standard deviation is not applicable, as the system transitions through these phases only once throughout the entire simulation.

Based on the findings outlined in Table \ref{results}, it's evident that the disparity in structure between phase I and the molecular liquid phase is minimal. Across the spectrum of states explored in this simulation, these two phases diverge primarily in temperature. Conversely, contrasting structural features emerge between the molecular liquid phase and the metallic liquid phase. In the metallic liquid phase, which is distinguished by higher pressures and temperatures relative to the molecular liquid, the average H-H distances exhibit a reduction. While the volumes of \ch{H6} remain approximately constant between the two states, the volume ratio is notably higher in the metallic liquid phase, indicating a denser atomic arrangement that manifests a more pronounced three-dimensional structure compared to the molecular liquid phase.

Noteworthy are the distinctions between phase I and phase IV*. As indicated in the table, hydrogen atom distances are shorter in phase IV*. Despite the system's exploration of states at temperatures akin to those associated with phase IV*, the higher pressure leads to reduced H-H distances (0.810 \AA\ in phase IV* versus 0.966 \AA\ in phase I). Moreover, the average volume in phase IV* is approximately halved compared to phase I. Although the reduction in average distances is not proportional, it suggests a planar configuration in phase IV*. The minimal standard deviation of distances hints at an approximately hexagonal structure, consistent with predictions from previous studies \cite{howie2012mixed, pickard2007structure}.

In phase V, the inter-atomic distances exhibit an average increase of 0.02 \AA\ compared to phase IV*. The significant surge in volume ratio and slight uptick in system volume imply a departure from hexagonal-planar symmetry in transitioning to phase V. Furthermore, at temperatures near those associated with phase IV* but at elevated pressures, only one point is linked with phase VI. Here, the \ch{H6} system is characterized by inter-atomic distances that are 0.058 \AA\ smaller than phase V, accompanied by a modest increase in volume and a substantial surge in volume ratio. Phase VI denotes a definite three-dimensional configuration for the system.

In phase V, the inter-atomic distances increase on average by 0.02 \AA\ compared to phase IV*. This transition is marked by a significant rise in the volume ratio and a slight increase in the system volume, suggesting a departure from hexagonal-planar symmetry as the system evolves into phase V. Furthermore, at temperatures close to those associated with phase IV* but at higher pressures, only one data point corresponds to phase VI. In this phase, the \ch{H6} system exhibits inter-atomic distances that are $0.058~\AA$ shorter than those in phase V, alongside a moderate increase in volume and a pronounced expansion in the volume ratio. Phase VI represents a distinct three-dimensional configuration for the system.

\section{Conclusions}
\label{sec:conclusion}
In summary, we identified an alternative method for achieving the high pressures and temperatures associated with a hydrogen system. By encapsulating six hydrogen atoms within the \ch{C20H20} dodecahedrane, we studied the evolution of the \ch{H6@C20H20} system. The \ch{C20H20} cage began from an initial configuration extended by a scaling factor of $k=1.2$, which initiated a compression corresponding to the symmetric radial vibration mode.

We explored the effect of this vibrational mode on the sequestered \ch{H6} system by analyzing geometric, energetic, and thermodynamic parameters over a total simulation time of 0.5 ps, using a time step of 0.5 fs. Correlations between the monitored variables showed that as the \ch{C20H20} cage compressed, the internal hydrogen atoms underwent structural changes. These changes were reflected in volume variations and led to significant dynamics in the HOMO-LUMO levels and the band gap of the system. Additionally, the compression increased the pressure of the encapsulated hydrogen system to hundreds of GPa. The temperature, calculated from the average kinetic energy of the internal hydrogen atoms, also spanned a wide range.

The \ch{H6} system explored different thermodynamic analogues states at each time step, covering a wide region in the $T-P$ diagram reported in the literature which includes high pressure hydrogen phases of interest: solid phases I, II, IV, V, and VI, as well as molecular and metallic liquid phases. This novel application of fullerane is significant for the field of high pressure hydrogen research and also in the hydrogen storage field. Further studies involving larger molecular cages or carbon nanotubes with more encapsulated atoms are necessary. This theoretical and computational study may pave the way for a new method of achieving dense hydrogen, potentially making the long-sought goal of room-temperature superconducting metallic hydrogen, first proposed by Wigner and Huntington nearly 90 years ago, a reality.

This study addresses a gap between experimental and theoretical approaches. While previous studies have introduced pressure and temperature as simulation parameters, our approach offers a new method for achieving those pressures and temperatures experimentally. This work not only advances our understanding of hydrogen behavior under extreme conditions but also suggests practical applications for creating metallic hydrogen.



\end{document}